\documentclass[prb,reprint,citeautoscript,floatfix]{revtex4-2}

\usepackage{amsmath}
\usepackage{amssymb}
\usepackage{graphicx}
\usepackage{hyperref}
\begin{document}
\title{Period tripling due to Josephson parametric down-conversion beyond the rotating-wave approximation}

\author{Lisa Arndt}
\email{lisa.arndt@rwth-aachen.de}
\affiliation{JARA Institute for Quantum Information, RWTH Aachen University, 52056 Aachen, Germany}

\author{Fabian Hassler}
\affiliation{JARA Institute for Quantum Information, RWTH Aachen University, 52056 Aachen, Germany}

\date{April 2022}
\begin{abstract}
Parametrically driven oscillators can display period-tripling in response to a drive at thrice the resonance frequency. In contrast to the parametric instability for period doubling, the symmetric fixed-point corresponding to the state of rest  remains stable at arbitrary strong driving for the tripling transition. Previously, it has been shown that fluctuations can circumvent this and induce a period-tripling instability. In this article, we explore an alternative way of inducing a period-tripling transition by investigating properties of period-tripling due to parametric down-conversion beyond the rotating-wave approximation. We show that despite the absence of an instability threshold, off-resonant frequency contributions can induce a period-tripling transition by activating the parametric down-conversion. Moreover, we study the subsequent period-tripled states of the Josephson potential and discuss the asymmetry between the clockwise and counter-clockwise rotating fixed-points that only arises beyond the rotating-wave approximation.
\end{abstract}
\maketitle

\section{Introduction}

In parametrically driven oscillators, period multiplication can be observed in the spontaneous emergence of phase-locked oscillations at an integer multiple of the driving period. The most commonly studied example is the period doubling in a degenerate parametric oscillator \cite{guckenheimer,strogatz:00,wustmann:19}. There, the system undergoes a pitchfork bifurcation as a function of the parametric driving strength. At the corresponding instability threshold, the symmetric state of rest turns unstable and is split-up into two continuously-emerging, symmetry-broken states \cite{strogatz:00,arndt:21}. This second-order instability distinguishes period doubling from multiple-period transitions, where the symmetric state of rest remains stable for any driving strength. However, thermal fluctuations  \cite{tadokoro:20} or quantum fluctuations \cite{arndt:22} can be employed to induce symmetry-breaking, multiple-period transitions despite the absence of a classical instability threshold. 

Following the transition from the state of rest, the dynamics of the system is subject to a discrete periodicity in phase space that emerges due to period multiplication. This periodicity can be used to engineer tunable energy band structures \cite{guo:13,guo:16,svensson:18,guo:20,lang:21}. In addition, the multiplicity of the states allows for the study of higher-order squeezing and multipartite entanglement \cite{braunstein:87,armour:13,chang:20} as well as multiple-state tunneling and correlations \cite{zhang:17,zhang:19,lorch:19,gosner:20,nathan:20}.
However, while period doubling systems have been analyzed in details even outside of rotating-wave approximations \cite{struble:63,starrett:74,garira:03,xu:07,navarrete:21}, studies of  multiple-period transitions have been focused, to our knowledge, solely on the system dynamics in the rotating frame.

In this article, we build on the rotating-frame results of Ref.~\cite{arndt:22} to investigated properties of the period-tripling transition and the subsequent period-tripled states beyond the rotating-wave approximation for Josephson parametric down-conversion. Circuit QED setups utilizing Josephson junctions as non-linear elements have been successfully employed in the past to observe multiple-period transitions \cite{denisenko:16,svensson:17,svensson:18,chang:20}. Here, we discuss the dynamics of a dc-biased Josephson junction coupled to a microwave resonator that implements the parametric drive by utilizing the ac-Josephson effect. For this setup, we discuss how off-resonant frequency contributions can be employed to induce a period-tripling transition. Additionally, we study the influence of the off-resonant effects on the characteristic 6-fold symmetry of the period-tripled states \cite{arndt:22}. We find that off-resonant contributions of the parametric drive lead to an asymmetry between the clockwise and counter-clockwise rotating fixed-points, even in the limit of small dissipation and weak detuning where the rotating-wave approximation is generally expected to be accurate.
\begin{figure}[tb]
	\centering
	\includegraphics{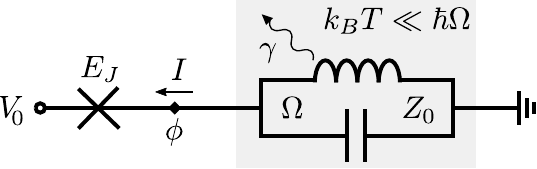}
	\caption{%
		The setup consists of a Josephson junction with Josephson energy $E_J$ biased by a dc-voltage $V_0$ and coupled to a microwave resonator. The resonator is  characterized by its resonance frequency $\Omega$ with a small bandwidth $\gamma$, as well as an impedance $Z_0$ at low frequency. The resonator is assumed to be coupled to a low temperature environment with $k_BT\ll \hbar \Omega$, such that thermal fluctuations are negligible. The voltage difference across the junction depends on the voltage $V=\hbar \dot\phi/2e$ across the resonator as well as the applied dc-voltage $V_0$. 
	}\label{fig:setup}
\end{figure}

The article is organized as follows. We introduce the microwave setup in Sec.~\ref{sec:setup}. In Sec.~\ref{sec:rwa}, we summarize the rotating-frame dynamics of the system based on Ref.~\cite{arndt:22}. We introduce the Poincare cross-section in Sec.~\ref{sec:poincare}, as a means to compare the rotating-frame results to the laboratory frame calculations. In Sec.~\ref{sec:tripling}, we discuss how off-resonant contributions of the drive can be employed to induce a period-tripling transition by ramping up the parametric driving strength. The arising period-tripled states are analyzed in Sec.~\ref{sec:states}, before we conclude in Sec.~\ref{sec:con}.

\section{Setup}\label{sec:setup}

Following Ref.~\cite{arndt:22}, we investigate a setup composed of a Josephson junction with Josephson energy $E_J$ that is biased by a dc-voltage $V_0$ in series with a microwave resonator. The resonator is characterized by its resonance frequency $\Omega$, a small bandwidth $\gamma$, as well as an impedance $Z_0$ at low frequency. Additionally, the resonator is assumed to be coupled to a low temperature environment with $k_BT\ll \hbar \Omega$, such that thermal fluctuations are negligible. The impedance $Z(\omega)$ of the resonator is given by
\begin{equation}\label{eq:impedance}
Z(\omega)=\frac{Z_0\Omega}{\gamma-i(\omega^2-\Omega^2)/2\omega}.
\end{equation}
The voltage difference across the junction is the difference of the voltage $V=\hbar \dot\phi/2e$ across the resonator and the applied dc-voltage $V_0$. To investigate period tripling, we choose the Josephson frequency $\Omega_J$ close to $3$ times the resonance frequency of the microwave resonator by setting the dc-bias voltage to $V_0=\hbar\Omega_J/2e$, with $\Omega_J=3(\Omega-\Delta)$. Here, the detuning $\Delta$ is assumed to be small with $|\Delta|\ll\Omega$. 

The impedance $Z_0$ determines the strength of the quantum fluctuations $(\delta\phi)^2= 4\pi G_QZ_0=\kappa/2$, with $G_Q=e^2/\pi\hbar$ the conductance quantum. The effects of quantum fluctuations on the period-tripling transition and the subsequent period-tripled states have been studied in Ref.~\cite{arndt:22} in the quasi-classical limit $\kappa\ll 1$ by employing a rotating-wave approximation. Here, we neglect the quantum fluctuations to instead focus on the classical description of the system in the laboratory frame and discuss effects that can only be studied outside of the rotating-wave approximation.

The classical system can be described by the second-order equation
\begin{equation}\label{eq:eqmotion}
\begin{aligned}
\ddot \phi&=-2\gamma \dot \phi-\Omega^2 \phi+16\Omega\epsilon\sin\left(\Omega_J t-\phi\right),
\end{aligned}
\end{equation}
with the driving strength $\epsilon=\kappa E_J/16 \hbar$. We introduce the dimensionless charge $q=\dot \phi/\Omega$ that relates to the current flowing through the resonator via $I=\hbar \dot q/4 e Z_0$.  For a better comparison to the rotating-frame results, we introduce the complex variable $\alpha=\phi+i q$ that satisfies the first-order differential equation
\begin{equation}\label{eq:eqmotionalpha}
\dot \alpha=-\gamma (\alpha-\overline{\alpha})-i\Omega\alpha+16i\epsilon\sin\left[\Omega_J t-\tfrac12\left(\alpha+\overline{\alpha}\right)\right].
\end{equation}

\section{rotating-wave approximation}\label{sec:rwa}
\begin{figure}[tb]
	\centering
	\includegraphics{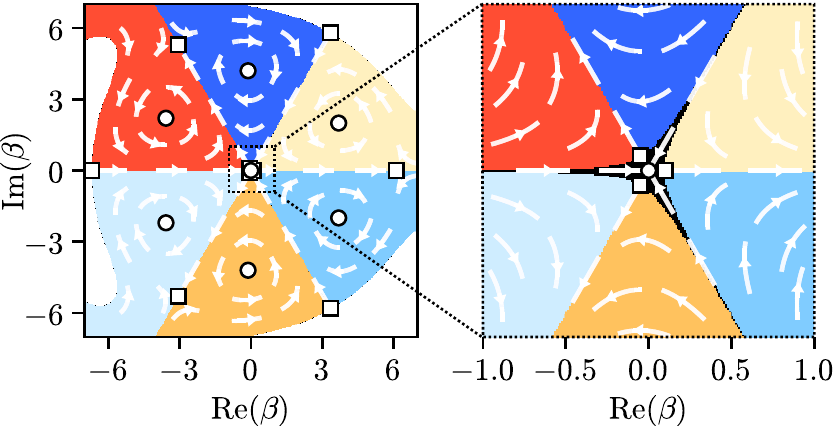}
	\caption{%
		(color online) Stability map and selected trajectories for the period-tripling system as described by the $3$-fold symmetric rotating-wave approximation in Eq.~\eqref{eq:rwamotion} for $\Delta=0$ and $\gamma=0.1\,\epsilon$. Within the shown region, the system exhibits $6$ stable (circles) and $6$ unstable (squares) fixed-points in the outer region as well as $3$ unstable and $1$ stable fixed-point in the center (zoomed-in on the right). The basins of attraction for each stable fixed-point are indicated by the colored regions. Blueish and reddish colors refer to clockwise and counter-clockwise rotation, respectively. The white region corresponds to higher-order fixed-points which are positioned outside of the displayed region and not of interest to the discussion in this article. 		%
	}\label{fig:rwa_stab}
\end{figure}

While this paper focuses on properties of the system beyond the rotating-wave approximation, for comparison, we first present the most important dynamical properties in the rotating frame. In the following, we give a short summary of the results found in Ref.~\cite{arndt:22}. 

In the limits $\gamma,\epsilon,|\Delta|\ll\Omega$, it is possible to perform a rotating-wave approximation \cite{armour:13} by introducing the slow, complex variable $\beta(t)$ via
\begin{equation}\label{eq:rwa}
\alpha(t)=\beta(t)e^{-i\Omega_J t/3},
\end{equation}
and neglecting all fast-oscillating terms. This leads to the autonomous differential equation \cite{lang:21}
\begin{equation}\label{eq:rwamotion}
\dot\beta=-\gamma \beta-i\Delta \beta-\epsilon \frac{\partial}{\partial \overline{\beta}}\left[\left(\beta^3-\overline{\beta}^3\right)f\left( \beta \overline{\beta}\right)\right],
\end{equation}
with $f(x)=16 J_3(x^{1/2})/x^{3/2}$ and $J_{3}$ the Bessel function of the first kind. The rotating-frame dynamics displays a $3$-fold rotational symmetry such that the transformations $\beta\mapsto \beta\, e^{2\pi i n/3}$, where $n\in \mathbb{Z}$, leave the equation of motion unchanged. 

A stability map of the rotating-frame dynamics in Eq.~\eqref{eq:rwamotion} is illustrated in Fig.~\ref{fig:rwa_stab}. An important feature of the stability map is the stable fixed-point at $\beta=0$ that corresponds to the state of rest of the oscillator which remains stable for any driving strength $\epsilon$ \cite{gosner:20}. However, in the limit of small dissipation and detuning $\gamma,|\Delta|\ll\epsilon$, the fixed point at the origin is closely surrounded by $3$ equidistant, unstable fixed-points that form a small, triangular basin of attraction around the state of rest \cite{arnold:89}. This is shown in the enlargement of Fig.~\ref{fig:rwa_stab}. Close to the origin, we can approximate $f(x)\approx 1/3$ to obtain the differential equation
\begin{equation}\label{eq:rwamotionvac}
\dot\beta=-\gamma \beta-i\Delta \beta+\epsilon\overline{\beta}^2.
\end{equation}
From this equation, we obtain the distance $|\beta|=(\Delta^2+\gamma^2)^{1/2}/\epsilon$ of the $3$ unstable fixed-points from the origin. In the following sections, we focus on the limit $\gamma,|\Delta|\ll\epsilon$, where the basin of attraction around the state of rest remains small. As we discuss in Sec.~\ref{sec:tripling}, this limit is favorable for inducing a period-tripling transition.

After an escape from the state of rest, \emph{e.g.}, due to thermal  \cite{tadokoro:20} or quantum fluctuations \cite{arndt:22}, the system ends up close to one of the $6$ outer stable fixed-points. Their dynamical properties can be investigated by expanding $f(x)\approx 1/3-x/48$ to the next order. In the limit $\gamma,|\Delta|\ll\epsilon$, all fixed points have the same amplitude $|\beta|=(48/5)^{1/2}$ and differ in phase by $\pi/3$ \cite{arndt:22}. Due to the additional mirror symmetry of Eq.~\eqref{eq:rwamotion} with respect to the real axis, the dynamics in the vicinity of the $6$ stable fixed-points alternates between clockwise and counter-clockwise rotation. In the limit $\Delta\to 0$, the system is equally likely to end up in any of the $6$ stable fixed-points, since each of the $3$ unstable fixed-points relevant for the period-tripling transition is equally connected to a mirrored pair of fixed-points in the outer region. For finite but small values of $\gamma/\epsilon$, the distance between these mirrored pairs of fixed-points decreases linearly with $\gamma/\epsilon$. 

At a fixed-point $\beta=|\beta|e^{i\varphi }$, the current in the laboratory frame approximately oscillates  as $I=\hbar \dot q/4 e Z_0=-(\hbar \Omega_J/12 e Z_0)|\beta|\cos[(\Omega_J t-3\varphi)/3]$. Since the full dynamics of Eq.~\eqref{eq:eqmotionalpha} remains invariant under the transformation $t\mapsto t+2\pi/\Omega_J$, only $3\varphi$ is a well-defined quantity in relation to the parametric drive. This is a characteristic property of period tripling and reflected by the $3$-fold symmetry in the rotating frame, where the system is invariant under the transformations $\varphi\mapsto \varphi +2\pi n/3$, with $n\in \mathbb{Z}$.

In the microwave setup discussed, the order parameter $3\varphi$ of the phase-locked, period-tripled states can be obtained by superposing $I^3$ with the original Josephson current $I_J\propto\sin(\Omega_Jt)$ or its time derivative $\dot I_J\propto\cos(\Omega_Jt)$. Averaging the resulting signal over times $t\gtrsim 6\pi/\Omega_J$ yields
\begin{equation}\label{eq:measure}
\langle I(t)^3I_J(t) \rangle\propto\sin(3\varphi),\quad \langle I(t)^3\dot I_J(t) \rangle\propto\cos(3\varphi),
\end{equation}
such that $3\varphi$ can be fully determined by these $2$ measurements.
\section{Poincare cross-section}\label{sec:poincare}
\begin{figure*}[th]
	\centering
	\includegraphics{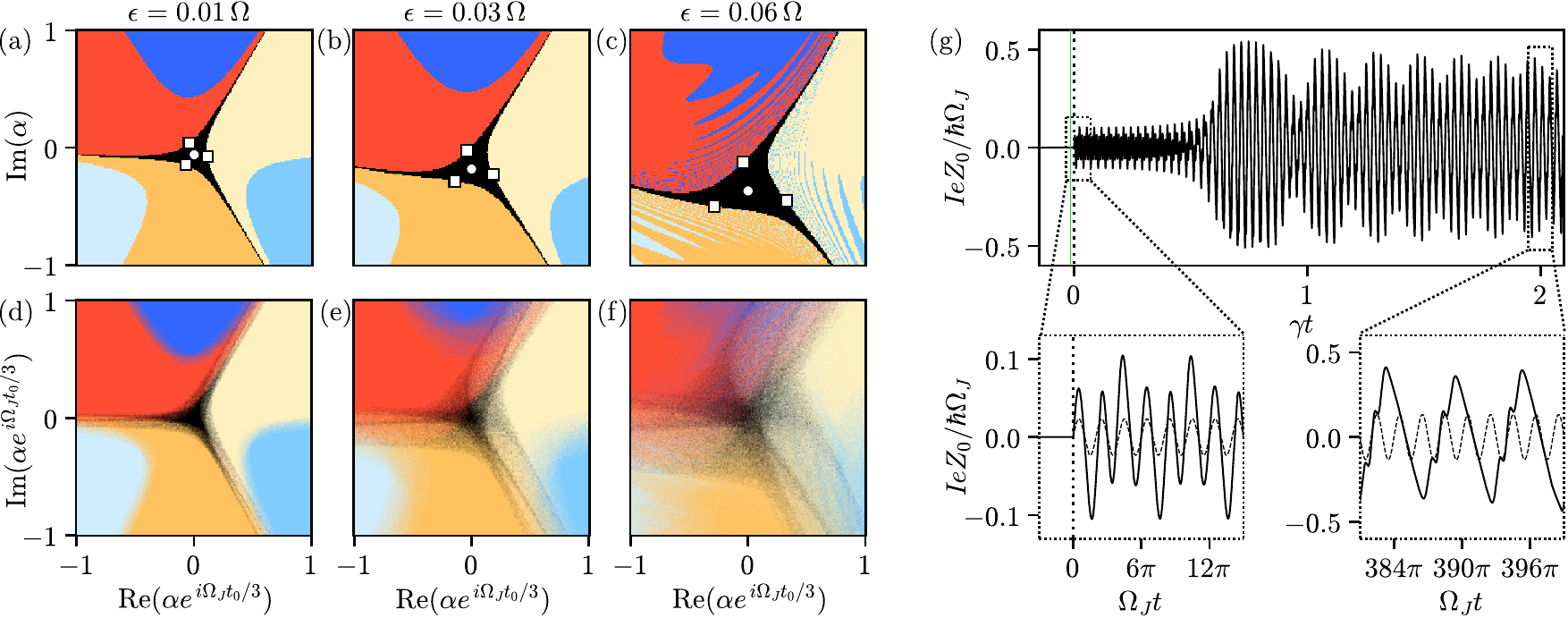}
	\caption{%
		(color online) (a)--(c) Poincare stability map of the dynamics in the laboratory frame [Eq.~\eqref{eq:eqmotionalpha}] at fixed $t_0=0$ for $\Delta=0$, $\gamma=0.1\,\epsilon$, and $\epsilon/\Omega=0.01,0.03,0.06$. The coloring is analogous to Fig.~\ref{fig:rwa_stab}. Note that for increasing $\epsilon/\Omega$, the displacement of the fixed-point, that corresponds to the state of rest in the rotating-frame, from the origin and its basin of attraction increases. (d)--(f) Corresponding averaged Poincare stability map for identical parameters. The map was obtained by averaging over $30$ trajectories at random, equally distributed values of the stroboscopic time $t_0$ for each point of the map. Since the rotated solution $\alpha(t) e^{i\Omega_J t_0/3}$ remains dependent on $t_0$, a statistical mixing between the different basins of attraction occurs in certain regions. This mixing is particularly strong close to the origin and at larger driving strength $\epsilon/\Omega$. (g) Exemplary current $I$ flowing through the junction during a period-tripling transition induced by abruptly switching on the parametric drive with $\epsilon=0.05\,\Omega$ at $t=0$ for $\Delta=0$ and $\gamma=0.1\,\epsilon$. The switch is indicated by the dotted line. The dashed line corresponds to the Josephson current $I_J\propto \sin(\Omega_J t)$ in arbitrary units. Following the switch, the system undergoes a transient response on the timescale $1/\gamma$ during which the current oscillates approximately with the driving frequency $\Omega_J$. As the system equilibrates, the current is phase-locked to the Josephson current and oscillates with the frequency $\Omega_J/3$.	%
	}\label{fig:tunneling_zoom}
\end{figure*}

The full system in the laboratory frame described by Eq.~\eqref{eq:eqmotionalpha} takes frequency contributions away from the resonance frequency of the oscillator into account. Therefore, the resulting signal can no longer be characterized by the single frequency $\Omega_J/3$. However, for sufficiently small $\gamma,\epsilon,|\Delta|\ll\Omega$, an analogous approach to Eq.~\eqref{eq:rwa} remains useful, \emph{e.g.}, to study how the stability map of Fig.~\ref{fig:rwa_stab} changes in the presence of the off-resonant frequency contributions. To this end, we define the Poincare map \cite{teschl:12}
\begin{equation}\label{eq:poincare}
\mathcal{P}\colon \alpha(t)\to\alpha(t+6 \pi/\Omega_J)
\end{equation}
that defines a stroboscopic evolution in time. Following the rotating-frame ansatz in Eq.~\eqref{eq:rwa}, the time steps $\delta t=6\pi/\Omega_J$ correspond to a full oscillation period of the period-tripled states. This choice ensures that in the limit $\gamma,\epsilon,|\Delta|\ll\Omega$ the points obtained by the Poincare map follow the rotating-frame evolution of $\beta(t)$. Similar to the rotating-frame description, the Poincare map of the full system displays fixed points characterized by $\mathcal{P}(\alpha)=\alpha$. However, it is worth noting that, conventionally, the Poincare map corresponds to a time evolution by $2\pi/\Omega_J$ in dependence on the periodicity of the time-dependent drive. As such, each fixed-point of the Poincare map defined in Eq.~\eqref{eq:poincare} that relates to a period-tripled state corresponds to a triplet of points in the conventional Poincare map. The points in this triplet have roughly the same amplitude and differ in phase by approximatly $2\pi/3$. This reflects the previously discussed property that only $3\varphi$ is a well-defined quantity for the period-tripled states.

The stroboscopic nature of the Poincare map requires a discussion of the origin in time $t_0$ at which the first point of the Poincare map is recorded \cite{bukov:15,zeuch:20}. For the rotating-frame results, a constant shift $t\mapsto t+t_0$ of the origin in time can be compensated by a simple rotation in phase space with $\alpha(t)\mapsto \alpha(t) e^{i\Omega_Jt_0/3} $. However, due to off-resonant frequency contributions, a constant shift of the origin in time translates not solely to a rotation in phase space for the full dynamics. As such, a Poincare map of $\alpha(t)$ taken at an initial time $t=0$ with $\alpha(0)=\alpha_0$ is not generally identical to a Poincare map of $\alpha(t)e^{i\Omega_Jt_0/3} $ taken at an initial time $t=t_0\neq0$ with $\alpha(t_0)e^{i\Omega_Jt_0/3}=\alpha_0$.

A Poincare stability map at fixed $t_0=0$ is displayed in Fig.~\ref{fig:tunneling_zoom}(a)--(c) for $\Delta=0$, $\gamma=0.1\,\epsilon$, and different values of $\epsilon/\Omega$. At small values of $\epsilon/\Omega$, the Poincare stability map closely resembles the stability map in the rotating frame displayed in Fig.~\ref{fig:rwa_stab}. In the case that the time-shift $t_0$ cannot be controlled experimentally, it is useful to discuss an averaged Poincare stability map that presumes uniformly distributed  time-shifts $t_0$, see Fig.~\ref{fig:tunneling_zoom}(d)--(f). Due to the remaining dependence of $\alpha(t) e^{i\Omega_J t_0/3}$ on $t_0$, a statistical mixing between the different basins of attraction occurs in certain regions. This mixing is particularly strong close to the origin and at larger driving strength $\epsilon/\Omega$.

\section{Period-tripling transition}\label{sec:tripling}

After introducing the formalism of the Poincare map in the previous section, we utilize it in this section to study the dynamics of the full system for small oscillation amplitudes. In the rotating-wave approximation in Eq.~\eqref{eq:rwamotionvac}, the lowest order contribution of the Josephson potential is quadratic in $\beta$ and the system displays a stable state of rest at the origin. For the full system, we can expand the Josephson term in Eq.~\eqref{eq:eqmotionalpha} to lowest order for small amplitudes $\alpha$ to obtain the differential equation of a driven harmonic oscillator
\begin{equation}\label{eq:eqmotionalphavac}
\dot \alpha=-\gamma (\alpha-\overline{\alpha})-i\Omega\alpha+16i\epsilon\sin\left(\Omega_J t\right),
\end{equation}
with the strongly off-resonant driving frequency $\Omega_J=3(\Omega-\Delta)$. Even though the coherent drive is strongly off-resonant, it dominates the dynamics for small amplitudes such that close to origin the system displays coherent oscillations at the frequency $\Omega_J\approx 3 \Omega$. In the limit $\gamma,\Delta\ll\Omega$, the amplitude of the oscillating current is given by $I_0= 9\hbar \epsilon/2 e Z_0$. For a Poincare map with time origin $t_0$, Eq.~\eqref{eq:eqmotionalphavac} results in the fixed point
\begin{equation}\label{eq:fixvac}
\alpha=\frac{2\epsilon}{\Omega}[ \sin(\Omega_J t_0)-3i\cos(\Omega_J t_0)].
\end{equation}
This corresponds to an average shift $|\alpha|\approx 4.3 \epsilon/\Omega$ of the fixed point from the origin and a maximum shift $|\alpha|= 6 \epsilon/\Omega$ for $\Omega_J t_0=n\pi$ with $n\in\mathbb{Z}$. As Fig.~\ref{fig:tunneling_zoom}(a)--(c) shows, these coherent oscillations
are not sufficient to destabilize the system and induce a period-tripling transition as the triangular basin of attraction is shifted in tandem with the fixed-point.

\begin{figure}[tb]
	\centering
	\includegraphics{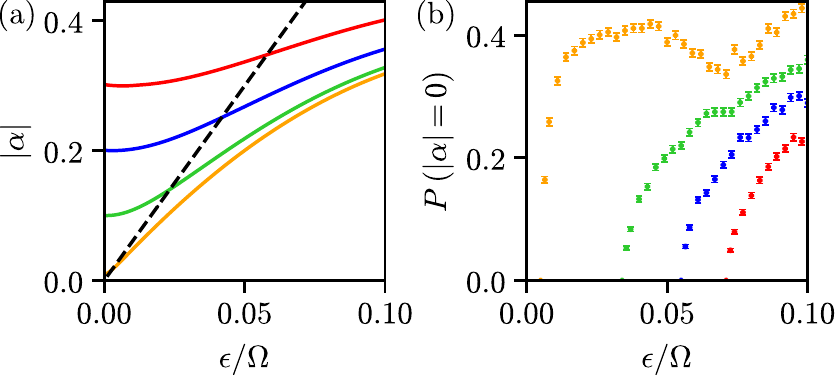}
	\caption{%
		(color online) (a) Numerical results for the minimum distance of the $3$ unstable fixed-points from the stable fixed-point at their center for $t_0=0$, $\Delta=0$, and $\gamma/\epsilon=0.01,0.1,0.2,0.3$ (solid lines from bottom to top). The maximum shift of the fixed point from the origin $|\alpha|\approx6\epsilon/\Omega$ for a time shift $t_0\approx 0$ is indicated by the black, dashed line. If this shift is smaller than the distance of the unstable fixed-points from the stable fixed-point at their center a ramp-up is unable to induce a period-tripling transition. (b) Probability that the origin at $|\alpha|=0$ lies outside of the basin of attraction of the central fixed point for a random value of $t_0$. The estimated probabilities and error bars were obtained by a simulation of Eq.~\eqref{eq:eqmotionalpha} for $\Delta=0$, $\gamma/\epsilon=0.01,0.1,0.2,0.3$ (from top to bottom), and $5000$ randomly chosen values of $t_0$ for each point.	%
	}\label{fig:tunneling_ana}
\end{figure}

However, the displacement of the fixed point and its basin of attraction from the origin proportional to $\epsilon/\Omega$ creates the opportunity to induce a period-tripling transition by ramping up the driving strength $\epsilon$. In the Poincare stability map, this ramp-up shifts the fixed point and its triangular basin of attraction outward. If the displacement of the fixed point is larger than the size of its basin of attraction after the ramp up, a period-tripling transition can be induced. As discussed in Sec.~\ref{sec:rwa}, the rotating frame approximation suggests a distance $|\beta|=\gamma/\epsilon$ of the $3$ unstable fixed-points from the state of rest for $\Delta=0$. For the full system, numerical results for the minimum distance of the unstable fixed-points from the stable fixed-point at their center are displayed in Fig.~\ref{fig:tunneling_ana}(a) for $\Delta=0$. These results show that a larger parametric driving strength $\epsilon/\Omega$ increases the size of the basin of attraction. While this effect is not advantageous for inducing a period-tripling transition, Fig.~\ref{fig:tunneling_ana}(a) also indicates that achieving a displacement larger than the size of the basin of attraction remains possible, especially for small ratios of $\gamma/\epsilon$.

Since the relaxation of the dynamics in the Poincare map takes place on a time scale $1/\gamma$, it is natural to assume that the time frame of the ramp-up has to be much smaller than $1/\gamma$. However, both the angle and the amplitude of the displacement of the stable fixed-point is strongly dependent on the time shift $t_0$ at which the Poincare map is taken. Since the size of the basin of attraction is only weakly dependent on the time origin $t_0$, the large displacement of the fixed point at $\Omega_J t_0=n\pi$ with $n\in\mathbb{Z}$ is favorable for inducing a period-tripling transition. In contrast, a time shift of $\Omega_J t_0\approx \frac{\pi}{2}+n\pi$ results in a basin of attraction that is larger than the displacement from the origin for the entire range of parameters studied in this article. As a result, the success of the ramp-up is dependent on the phase of the parametric drive at the time of the ramp up. Therefore, a successful period-tripling transition requires a nearly instantaneous ramp up on a time scale $\ll 2\pi/\Omega_J$ which in turn requires a large accessible bandwidth. Theoretically, an ideal ramp-up can be approximated by multiplying the Josephson potential in Eq.~\eqref{eq:eqmotionalpha} with the step function $\Theta(t)$ such that the parametric drive is turned on when the Josephson current crosses through zero. The resulting period-tripling transition is displayed in Fig.~\ref{fig:tunneling_zoom}(g). Following the switch, the system undergoes a transient response on the timescale $1/\gamma$ during which the current oscillates approximately with the driving frequency $\Omega_J$. As the system equilibrates, the current is phase-locked to the Josephson current and oscillates with the frequency $\Omega_J/3$.
We note in passing that without control over the time of the ramp-up of the amplitude $\epsilon$, the success of the switch is probabilistic. Numerical predictions for the probabilities of success for a random distribution of $t_0$ are displayed in Fig.~\ref{fig:tunneling_ana}(b). Even at the very small dissipation rate $\gamma=0.01\,\epsilon$, the probability of success does not exceed $50\%$. 

\section{Period-tripled states} \label{sec:states}

\begin{figure*}[t]
	\centering
	\includegraphics{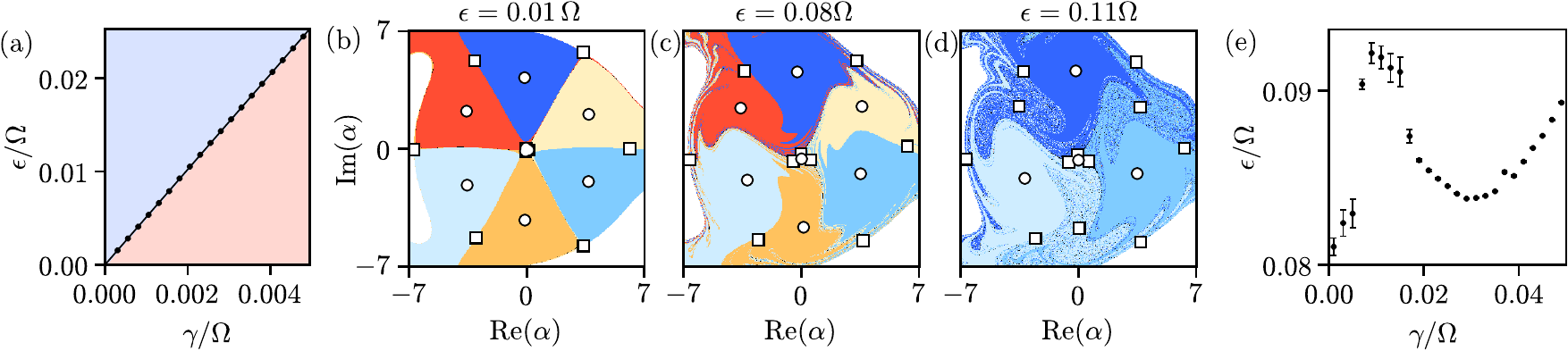}
	\caption{%
		(color online) (a) Driving strength $\epsilon/\Omega$ at which the unstable Poincare fixed-points surrounding the central fixed-point switch from bordering clockwise rotating regions to counter-clockwise rotating regions in the stability map. This property directly indicates which fixed-point is dominantly accessed after the transition from the central fixed-point. The points were obtained by numerical simulation of Eq.~\eqref{eq:eqmotionalpha} for $\Delta=0$. The line indicates a linear fit that returns an inverse slope of $\gamma/\epsilon=0.20$. (b)--(d) Poincare stability map of the laboratory frame dynamics in Eq.~\eqref{eq:eqmotionalpha} at fixed $t_0=0$ for $\Delta=0$, $\gamma=0.1\,\epsilon$, and $\epsilon/\Omega=0.01,0.08,0.11$. The coloring is analogous to Fig.~\ref{fig:rwa_stab}. At $\epsilon/\Omega\approx 0.08$--$0.09$, the counter-clockwise rotating fixed-points (reddish colors) turn unstable. (e) Upper estimate for the critical driving strength at which the counter-clockwise fixed-points become unstable for $\Delta=0$. The results were obtained by numerical simulations of Eq.~\eqref{eq:eqmotionalpha} for $10$ random initial points $\alpha_0$ on a circle of radius $0.1$ around the counter-clockwise fixed-points. The error bars were estimated by the distribution of the lowest driving strength at which the initial points lead to a dynamics in the vicinity of a clockwise fixed-point at time $\tau=1000/\gamma$. As the change in stability is connected to the emergence of a long time-scale over which the counter-clockwise fixed-points initially appear to be stable, the results should be considered as an upper bound of the critical driving strength.         	%
	}\label{fig:mixing}
\end{figure*}

In this section, we take a closer look at the period-tripled states that emerge following the period-tripling transition. Within the framework of the rotating-wave approximation, the system ends up close to one of the $6$ outer stable fixed-points after an escape from the state of rest, see Sec.~\ref{sec:rwa}. For $\Delta=0$, the rotating-wave approximation predicts that the system is equally likely to end up in any of the $6$ stable fixed-points, since each of the $3$ unstable fixed-points relevant for the period-tripling transition is equally connected to a mirrored pair of stable fixed-points in the outer region. This no longer holds for the corresponding $6$ Poincare fixed-points of Eq.~\eqref{eq:eqmotionalpha}. Instead, either the clockwise or the counter-clockwise fixed-point of the mirrored pair dominates the dynamics after the period-tripling transition. This is evident in the stability map in Fig.~\ref{fig:tunneling_zoom}(a). 

Whether the system at $\Delta=0$ is more likely to end up close to a clockwise or a counter-clockwise rotating point depends on the ratio $\gamma/\epsilon$. Figure~\ref{fig:mixing}(a) displays the parameter regimes for which the unstable Poincare fixed-points surrounding the central fixed-point are bordered by either clockwise or counter-clockwise rotating regions in the stability map. This directly implies which of the fixed-points is dominantly reached after the transition. The two different regimes are separated by a line at $\gamma\approx 0.2\,\epsilon$. At lower dissipation rates, the system is more likely to end up in a counter-clockwise rotating region, while for higher dissipation a clockwise rotation is more likely.

However, we observe this distinction only for small values of $\epsilon/\Omega\lesssim 0.03$. In this regime, the borders between different areas of the stability map remain smooth, as shown in Fig.~\ref{fig:mixing}(b). At larger driving strength, the basins of attraction for different fixed-points begin to mix as indicated in Fig.~\ref{fig:mixing}(c). This mixing is strongest in the vicinity of the unstable fixed-points, rendering a clear distinction between the clockwise and counter-clockwise regime impossible.

We observe that, at an even larger value $\epsilon/\Omega =  0.11$ of the driving strength, the counter-clockwise (reddish) solution disappears altogether, see Fig.~\ref{fig:mixing}(d).   We associate this transition to the fact that the counter-clockwise fixed-points turn unstable altogether. This is based on the numerical evidence
that this switch happens very suddenly at a certain threshold in driving strength and that the switch is not preceded by a strong shrinking of the counter-clockwise basins of attraction.  Figure~\ref{fig:mixing}(d) shows that the basins of attraction of the counter-clockwise fixed-points before the transition are largely incorporated into the basins of attraction of their clockwise counterpart in the mirrored pairing described above. The change in stability is connected to the emergence of a long time scale over which the counter-clockwise fixed-points initially appear to be stable before `escaping' to the clockwise solution. The long-time scale makes it difficult to determine the critical driving strength numerically very accurately.  Figure~\ref{fig:mixing}(e) shows an upper bound for the critical driving strength that was obtained for a simulation time $\tau=1000/\gamma$. The numerical results indicate that the critical driving strength is only weakly dependent on $\gamma/\Omega$ in the  parameter regime considered. The critical driving strength is given by  $\epsilon/\Omega\approx 0.08$--$0.09$. 

\section{Conclusion}\label{sec:con}

In this manuscript, we have investigated properties of the period-tripling transition and the subsequent period-tripled states of the Josephson potential beyond the rotating-wave approximation. After a short summary of the previously well studied rotating-frame dynamics, we discussed the properties of the state of rest in the laboratory frame. Here, we have shown that off-resonant effects can be employed to induce a period-tripling transition by ramping up the driving strength $\epsilon$ over a short time $\ll2\pi/\Omega_J$. This constitutes a classical alternative to the escape in the presence of quantum fluctuations previously studied in Ref.~\cite{arndt:22}. Furthermore, we have discussed that either the clockwise or the counter-clockwise period-tripled states of the system dominate after the escape from the state of rest. The more likely rotation depends on whether $\gamma$ is larger or smaller than $0.2\,\epsilon$. This result is in contrast to the rotating-wave approximation, which, for $\Delta=0$, predicts an equal likelihood to end up in any of the $6$ stable fixed-points. Lastly, we have investigated the critical driving strength $\epsilon/\Omega\approx 0.08$--$0.09$ at which the counter-clockwise Poincare fixed-points in the laboratory frame turn unstable altogether. Our results highlight the necessity of critically examining the results of the rotating-wave approximation. In particular, it is a-priori difficult to predict which of the results in the rotating-frame remain valid in a more accurate theoretical description of the system.\\
\begin{acknowledgments}

We are grateful to E. Varvelis for helpful discussions. This work was supported by the Deutsche Forschungsgemeinschaft (DFG) under Grant No.~HA 7084/6-1.

\end{acknowledgments}

\end{document}